# Review of foundational concepts and emerging directions in metamaterial research: Design, phenomena, and applications

Jade E. Holliman Jr, H. Todd Schaef, B. Peter McGrail, and Quin R.S. Miller *

Pacific Northwest National Laboratory, Richland, Washington 99354, USA

ABSTRACT

In the past two decades, artificial structures known as metamaterials have been found to exhibit extraordinary material properties that enable the unprecedented manipulation of electromagnetic waves, elastic waves, molecules, and particles. Phenomena such as negative refraction, bandgaps, near perfect wave absorption, wave focusing, negative Poisson's ratio, negative thermal conductivity, etc., all are possible with these materials. Metamaterials were originally theorized and fabricated in electrodynamics, but research into their applications has expanded into acoustics, thermodynamics, seismology, classical mechanics, and mass transport. In this Research Update we summarize the history, current state of progress, and emerging directions of metamaterials by field, focusing the unifying principles at the foundation of each discipline. We discuss the different designs and mechanisms behind metamaterials as well as the governing equations and effective material parameters for each field. Also, current and potential applications for metamaterials are discussed. Finally, we provide an outlook on future progress in the emerging field of metamaterials.

## I. INTRODUCTION

Metamaterials are artificial materials composed of unit cells that work collectively to produce unusual, unique physical properties not found in natural materials or traditional composite materials. In turn, these properties enable metamaterials to manipulate propagating waves as well as the transport of matter, thus opening the door to a world of new devices with extraordinary abilities and potential applications. Over the years several definitions of metamaterials have been provided in literature, ranging from broad to narrow. The broader view is that a metamaterial is composed of artificial unit cells or atoms that act as one to produce properties not seen in natural materials[1-3]. More narrow definitions include but are not limited to ones that emphasize that effective material parameters are present[4] or that emphasize mathematical properties[5] such as linearity. Generally, the unit cells are arranged in a periodic array that is a key component to most metamaterials; however, this is not a requirement for all metamaterials. Depending on the operating wavelength, metamaterials can be constructed to be as small as nanoscale, with units of atomic lattices, to as large as meter-scale with units composed of strategically placed resonators. Over the past two decades many advancements have been made in the metamaterial field, but the idea was originally conceived in 1967 by physicist Victor Veselago[6].

Veselago wondered what if the electromagnetic parameters electric permittivity ($\varepsilon$) and magnetic permeability ($\mu$) in Maxwell's equations were simultaneously negative. He then realized that one would still get a propagating wave, however, it would be backwards meaning it has antiparallel wave vector and Poynting vectors[7]. Natural material with these parameters does not exist so Veselago further hypothesized possibilities for materials having such parameters. Veselago called these materials "left-handed substances" because when $\varepsilon$ and $\mu$ are simultaneously negative, electric field vector E, magnetic field vector H, and wave vector k form a left-handed set of vectors. . They are also known as double negative (DNG) materials. He proposed that when a material has simultaneous negative $\varepsilon$ and $\mu$, an electromagnetic wave incident on the material will exhibit negative phase velocity and the material will have a negative index of refraction, a reversal of Snell's law. He also proposed that the Doppler effect and Cherenkov radiation would be reversed. In the late 1990s J.B. Pendry and colleagues began preliminary work in the steps to realizing this left-handed material. Pendry et al.[8] theorized and computationally verified a cubic lattice structure made a of extremely thin wires capable of effective plasma frequency in the GHz regime with $\varepsilon$ of negative one Experimental verification of such a structure came soon after[9]. .Pendry et al.[10] then achieved negative permeability using very thin, periodically arranged nonmagnetic conducting sheets that formed cylindrical microstructures known as split ring resonators (SRRs). From these discoveries, D.R. Smith et al.[11] created a composite material consisting of periodically placed thin wires and SRRs that exhibited simultaneous negative permittivity and permeability. A negative index of refraction was experimentally proven at microwave frequencies the next year by R.A Shelby et al.[12], just as Veselago predicted.

Metamaterials are not the only artificial materials created to manipulate wave propagation. Photonic crystals (PTCs) [13, 14], artificial crystals comprised of periodically varying dielectric lattice structures with high index contrast, can manipulate electromagnetic waves and create band gaps. The idea of PTCs originates from two separate papers published in 1987 by Eli Yablonovitch[15] and Sajeev John[16]. They proposed material designs that affect photons in the same way semiconductor crystals affect electrons and create an electromagnetic bandgap. Although their approaches were different in that Yablonovitch intended to control spontaneous emission while John wanted to create strong photon localization, both researchers wanted to find the photonic bandgap. PTCs can behave as metamaterials[17] by exhibiting effective propagation properties[18, 19] or possessing effective parameters[20, 21] but, they are generally classified separately. There is one key difference that sets them apart; PTCs and other artificial crystals (i.e., phononic crystals [PNCs][22, 23] and platonic crystals [PLTCs][24]) can only manipulate waves that are of comparable size to their lattice constant, which typically results in a lattice constant about half the size of the operating wavelength. However, metamaterials can manipulate waves of the same size as their lattice constant or larger, giving metamaterials a much wider range of application. Metamaterials tend to employ geometry and resonance, among other mechanisms, to help achieve their novel properties. For example, electromagnetic metamaterials (EMs) often use magnetic resonance[25-28] while a class of metamaterials known as hyperbolic metamaterials[29-31] utilize high anisotropy instead of resonance.

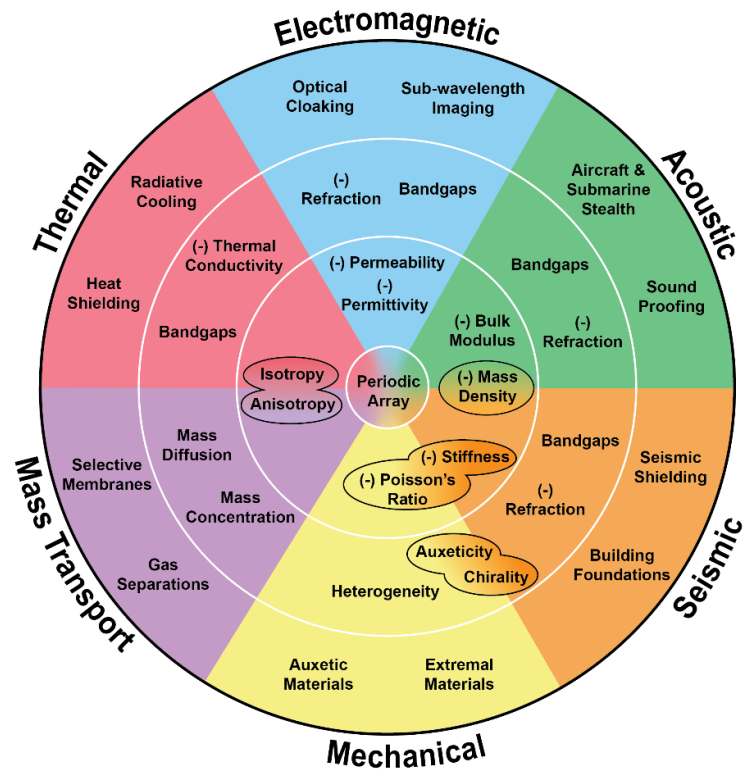

**Figure 1.** Relationships between parameters, phenomena, and applications for six types of metamaterials. The center of the figure contains some parameters of metamaterials, the middle ring contains related phenomena, and the outer ring contains practical applications. Most metamaterials implement periodic arrays into their structures.

Metamaterial research began in the electromagnetic field but has expanded to cover many physical domains (i.e., acoustics, thermodynamics, classical mechanics, seismology, and mass transport). Following the success of EMs researchers wondered if the metamaterial concept could be applied to other waves, leading to the first acoustic metamaterial in 2000[32]. Several years later coordinate invariant mathematical methods for creating EMs were developed opening the door for other physical domains possessing coordinate invariant governing equations. This realization lead to thermal metamaterials in 2008[33, 34], seismic metamaterials in 2012[35, 36], and mass transport metamaterials in 2016[37, 38]. Mechanical metamaterials do not have as clear of an origin, but materials classified as mechanical metamaterials today have been around for many years[39, 40]. Due to the vast amount of information on metamaterials, we provide review articles for each category, excluding mass transport. In this Research Update, we identify the established and emerging metamaterial fields. We discuss the different designs and mechanisms behind metamaterials as well as the governing equations and effective material parameters in each field. We will discuss the many interesting phenomena that occur such as negative refraction, bandgaps, wave attenuation, wave absorption, wave focusing, negative Poisson's ratio, negative stiffness, etc. Over the past two decades, the metamaterial field has grown significantly so we divide it into six fundamental categories: 1) electromagnetic, 2) acoustic, 3) thermal, 4) seismic, 5) mechanical, and 6) mass transport. **Figure 1** shows these six categories in a circular diagram that is read from the center outward. The center of the circle contains effective parameters that help produce the physical phenomena in the second, middle section. The third, outer section shows current and potential applications. The parameters and phenomena with blended colors are shared between two or more categories. Furthermore, this review summarizes the current progress and highlights existing work in each area.

## II. ELECTROMAGNETIC METAMATERIALS

EMs operate at several different frequencies depending on the operating wavelength and target application. Thus, EMs are comprised of several types of metamaterials that apply to specific wavelengths including radio waves[41], microwaves[42], terahertz[43, 44], infrared[45], and visible light (i.e., photonic metamaterials)[46, 47]. EMs are made possible through the manipulation of Maxwell's equations (**Table 1**). The equations describe how electric and magnetic fields propagate and interact with one another and with different mediums; therefore, being able to manipulate these equations subsequently allows electromagnetic waves also to be manipulated. Two fundamental parameters within these equations determine the electromagnetic properties of a medium: 1) electric permittivity (ε) and 2) magnetic permeability (μ). These properties are the effective parameters of EMs and give rise to their unique properties. Most materials have positive permittivity and permeability, a few have negative permittivity and positive permeability or vice versa, but no natural material has negative permittivity and permeability simultaneously. As theorized and later proven, a material that possesses negative permittivity and permeability simultaneously has a negative index of refraction. In nature, all mediums have a positive index of refraction in which incident electromagnetic waves are refracted at a positive angle to the normal. A negative index of refraction means that incident waves are refracted at a negative angle to the normal, in other words, on the same side of the normal as the incident wave. These materials are known as negative index materials (NIMs) and can serve as metamaterials by themselves or can be combined with other materials to form a metamaterial. In 2000, Pendry et al.[48] gave practical use to NIMs by proposing that a slab of NIM can be a perfect lens that can produce near-field subwavelength images beyond the diffraction to which traditional optical lenses are limited. The diffraction limit does not allow details that are smaller than half the wavelength of light to be resolved. These extremely fine details are carried by evanescent waves that exponentially decay in space. However, in theory, a perfect lens made of a NIM can amplify evanescent waves and focus them with the propagating waves, creating an image with greater detail than one from an optical lens. A schematic drawing of the perfect lens is shown in **Figure 2**. The perfect lens concept was proven a few years later[49, 50]; however, evanescent waves were observed to decay again at a certain material thickness, calling for improved lens designs. Also, these lenses are referred to as superlenses and not perfect lenses because of the inherent energy loss associated with NIMs.

Superlenses have since been experimentally demonstrated at optical[51, 52], microwave[53], and infrared frequencies[54-57] with different designs. These lenses are capable of imaging as small as 60 nm, but high energy losses still significantly reduce their efficiency. Because of the design of the slab of NIM superlens, it is limited to near-field applications. To bring the subwavelength image into the far-field, far-field superlenses have been proposed and experimentally achieved[58-60]. The far-field lens enhances and then converts evanescent waves into propagating waves, thus allowing the details they carry to be seen in the far-field. Hyperlenses in the far-field[61-64] also are capable of subwavelength imaging. A hyperlens is an alternative to a superlens that is anisotropic with hyperbolic dispersion. Lenses[65-67] also can be created from the transformation method discussed below. More information on the physics and types of metamaterial lenses is given in the reviews[68, 69].

Perhaps more interesting than a lens with subwavelength capability is the electromagnetic cloak. An electromagnetic cloak redirects incident electromagnetic waves around an object and recombines them on the other side in the same direction they entered, as if the object is not there (**Figure 2, Figure 3a**). At optical frequencies, this is perceived as invisibility to an observer. Such a device was theoretically conceived using coordinate transformations by U. Leonhardt[70] and Pendry et al[71]. Maxwell's equations are invariant under coordinate transformation meaning the equations remain true after being transformed to another coordinate system. This is the foundation of the transformation method known as transformation optics[72] for light waves. This method began in optics as a strategy to create EMs that allows any deformation of an electromagnetic field to be physically created with the corresponding spatial distribution of material based on the coordinate transformation. The resulting material usually is highly anisotropic, spatially complex, and has varying indices of refraction, which makes them extremely hard to fabricate. A microwave cloak consisting of SRRs is one of the earliest experimentally achieved cloaks[73]. The cloak is created by squeezing space from an arbitrary volume into an annular shell leaving an opening in the center. The shell is the transformed space made to redirect the incident light and surrounds the object being concealed. A cloak also can be created by squeezing the space into either a line[74] or a sheet[75, 76]. Based on the coordinate transformation, required electromagnetic properties are determined, and the cloak is created. Optical cloaks able to cloak objects on the micrometer scale have been experimentally achieved using transformation optics[77-79]. Electromagnetic cloaks have come a long way but require more research for practical applications. In addition to transformation optics, there are other methods for creating cloaks and EMs in general. Many involve implementing metasurfaces[80-82] that do not rely on the bulk constitutive properties of a composite material. A metasurface is an extremely thin material, smaller than the wavelength, made up of periodic units capable of manipulating wavefronts. A popular design technique known as inverse design[83-85] aims to achieve optimal metamaterial structure based on the desired characteristics. This is done using numerical optimizations (i.e., shape optimization, topology optimization, and genetic algorithms) and numerical analyses (i.e, finite element method and finite-difference time-domain) which solve Maxwell's equations providing the parameters required to fabricate the desired metamaterial and provide computational simulations of the resulting metamaterial. Antennas transmit and receive electromagnetic waves and have been used in televisions, telephones, radios, etc., for years. The range and efficiency of antennas always have been limited; however, with metamaterials, their capability and performance can be improved. Metamaterial antennas are comprised solely of metamaterials, or metamaterial components and techniques are implemented (**Figure 2**). In general, metamaterial antennas can be broken down into four categories: 1) metamaterial loaded, 2) composite right/left handed (CRLH)-based, 3) metasurface loaded, and 4) metamaterial inspired. Metamaterial loading involves loading the antenna with effective media that interact with the antenna to produce the desired metamaterial properties. Effective media include resonators, PTCs, negative permittivity/permeability material, or extreme permittivity/permeability material. CRLH-based antennas implement left- and right-handed material properties depending on the frequency. They are created through transmission line and resonator methods and often use a CRLH unit cell to produce the desired properties. Metasurface loading involves loading the antenna with metasurfaces with properties such as bandgaps, high impedance surfaces, and reactive impedance surfaces[86-88]. Metamaterial inspired antennas use one or more than two unit cells arranged in a particular array to produce the desired metamaterial properties[89-91]. These unit cells can be SSRs or can be engraved into the actual antenna surface[92]. All design methods have been experimentally proven to miniaturize antennas and improve gain and bandwidth. Electromagnetic wave absorption and sensing is another important application of EMs. Absorbers can be used for radar echo reduction, reducing unwanted radiation in antennas, protecting people from harmful radiation in medical devices, and reducing any other electromagnetic interference[93]. There are two types of absorbers: 1) resonant and 2) broadband. Resonant absorbers use resonance to interact with the incident waves at a certain frequency, while broadband absorbers have properties that are frequency independent, allowing them to operate over a wider bandwidth. Perfect absorption that allows no scattering or reflection can be obtained through metamaterial absorbers. Landy et al.[94] developed a perfect electromagnetic wave absorber in the microwave frequency range that was able to absorb nearly 100% of the incident wave. Soon metamaterial absorbers were expanded to other frequency ranges including terahertz[95-97], infrared[98, 99], and visible light[100]. More recently, W. Jiang et al.[101] used three-dimensional printing to develop a metamaterial absorber that is capable of more than 90% absorption in the radio wave frequency. Metamaterial sensors[102] mainly are used for material characterization[103, 104] and non-destructive evaluation[105]. Generally, sensors must absorb some of the incident waves to detect them, so metamaterial absorbers and sensors are often one in the same[106, 107]. They can detect extremely small chemical and biological attributes or changes as well as pressure changes. They commonly are applied in microwave and terahertz frequencies [108] but also have been tested for other frequencies.

All the metamaterials mentioned so far are classified as passive and static meaning that once fabricated, the metamaterial cannot be altered and possesses a narrow field of operation. Active metamaterials[109] are a class of metamaterials with functionalities that can be actively tuned or switched. They can be tuned in real time to fit a changing environment and respond accordingly. There are two general classes of active metamaterials: 1) mechanical reconfiguration materials and 2) active materials. Mechanical reconfiguration refers to altering structural qualities such as lattice constants, resonator shapes, or spatial arrangements using external stimuli that change the response of the metamaterial. Commonly used external stimuli include heat[110, 111], electricity[112], and light[113-115]. Alternatively, active materials can be implemented into the structure of metamaterials to create an active metamaterial. Active materials are sensitive to external stimuli, thus enabling their active control. Active materials include devices like varactor diodes[116-118], semiconductors[119, 120], liquid crystals[121, 122], phase change materials[123, 124], etc. Active metamaterials are seen across physical domains so others will be discussed in their respective sections.

### III. ACOUSTIC METAMATERIALS

Advances in metamaterials caused researchers to wonder if the theory of EMs can be applied to other fields. Soon enough, these methods were carried over to the acoustic field and applied to sound waves. This started a whole new class of metamaterials called acoustic metamaterials (AMs)[1, 125]. Like EMs, AMs have effective parameters that give rise to their extreme properties. Those properties are the bulk modulus ($\kappa$) and the mass density ($\rho$), which are mathematically analogous to permittivity and permeability, respectively. Negative bulk modulus and/or negative mass density gives rise to negative refraction and other unnatural effects. The acoustic wave equation (**Table 1**) is the governing equation for acoustic wave propagation and contains the effective parameters. Negative refraction can be used to create an acoustic band gap like the electromagnetic band gap seen in EMs. PNCs also were created to manipulate sound waves in the same way that PTCs manipulate electromagnetic waves. Phonons are units of vibrational energy analogous to photons. The first PNCs were introduced in the early 1990s by Sigalas and Economou[126] and Kushwaha et al.[127]. Phononic crystals achieve sound wave control through Bragg scattering while many AMs use a mechanism known as local resonance to control wave propagation. Local resonance refers to wave coupling that occurs when resonators are arranged periodically with oscillators or other resonators. In addition, Helmholtz resonance, which also is used in AMs, refers to the resonance phenomenon of air in a cavity. Acoustic metamaterials and PNCs are similar, but the range of frequencies effected by PNCs is limited by anisotropy and the lattice constant of the crystal while the range of AMs is unrestricted. AMs can manipulate waves of a much larger wavelength than their lattice size, allowing a wider range of application. The limitation of PNCs is due to their use of Bragg scattering that only produces a band gap of the same order as the lattice constant.

Acoustic metamaterials originally focused on sound attenuation, which was achieved by Z. Y. Liu et al.[32] in 2000. Their AM was sonic crystals comprised of centimeter-sized lead spheres coated in rubber and arranged in a square lattice with a lattice constant of 1.5 cm. This device uses local resonance to produce negative elastic constants, subsequently causing attenuation. In addition to attenuation AMs are capable of cloaking, subwavelength imaging, absorption, and sensing and are engineered at macroscale and microscale levels. Macroscale AMs involve periodic structures or transformation acoustics while microscale AMs involve the manipulation of phonons. Acoustic attenuation often is achieved through mechanisms such as Bragg scattering and local resonance. Membrane-type AMs have been shown to implement these mechanisms and achieve high attenuation for low-frequency sound (<500 Hz)[132, 133, 139, 140]. In addition, there are groups working on active acoustic metamaterials with some promising applications for sound attenuation[141, 142]. For example, Xia et al.[141] theorized a thermally tunable acoustic metamaterial for underwater sound blocking application.

The function of the acoustic cloak is similar to that of electromagnetic cloak, but instead of visual invisibility, audible invisibility is achieved. Using transformation acoustics, it was theorized in two dimensions[143] and then in three dimensions[144] due to the realization of the form-invariant linear acoustic equations. Many cloaks implement the traditional transformation method in which there is an annular shell structure that protects objects within the shell from outside sound waves by redirecting the waves around the shell. The waves then continue along their original propagation path effectively exhibiting acoustic invisibility. Many groups have fabricated functioning acoustic cloaks[145-148]; however, practical applications have yet to be realized because of the cloak's ability to perform only at small scale and the fabricating the cloaks is challenging because of their properties. Metasurface acoustic cloaks[149-151] are a promising alternative to transformation-based cloaks. These cloaks use very little material with less extreme properties. The acoustic lens (**Figure 2**), like the electromagnetic lens, is used for subwavelength imaging beyond the diffraction limit.

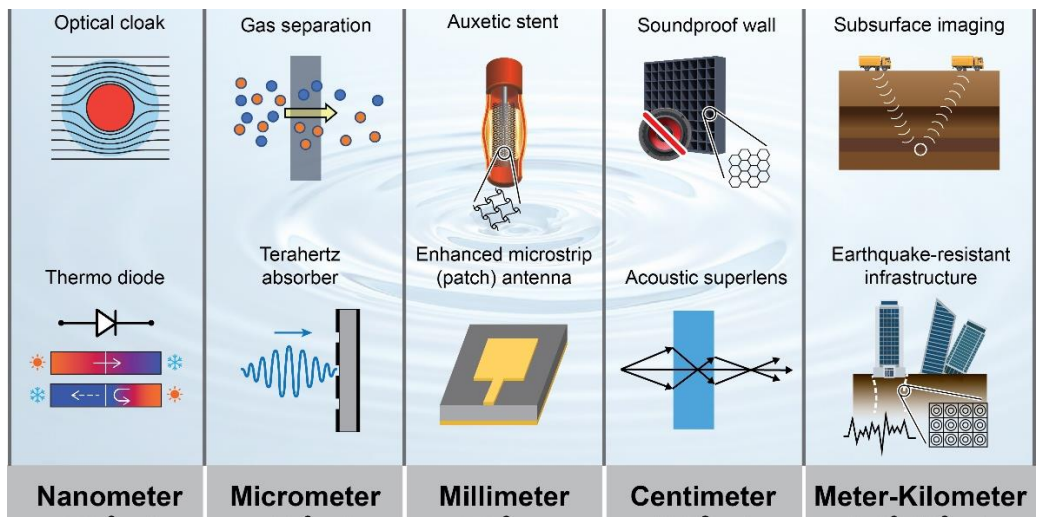

**Figure 2.** Examples of different characteristic length scales for metamaterial phenomena and applications. Metamaterials are fabricated in varying sizes depending on the operating wavelength and their intended function. Optical cloaks[77, 79] and thermal diodes[128, 129] are constructed on a nanometer scale to manipulate particles such as photons and phonons. Gas separation membranes[37, 130] and terahertz absorbers[96, 97] operate on a micrometer scale to manipulate molecules and high frequency terahertz waves. Auxetic stents[131] and microstrip (patch) antennas[89, 92] are constructed with millimeter dimensions to manipulate mechanical motion and electromagnetic waves. The soundproof wall[132, 133] and acoustic superlens[134, 135] have periodic structures on the centimeter scale to manipulate sound waves. Subsurface imaging[136] and earthquake-resistant infrastructure[137, 138] are constructed on the meter to kilometer scale to manipulate seismic waves.

Near-field superlenses[134, 135, 152] amplify evanescent waves while far-field hyperlenses[153, 154] convert evanescent waves to propagating waves. Other far-field lenses that do not use high anisotropy have also been designed[155, 156]. Acoustic lenses show great promise for applications such as medical imaging or building structure imaging. For example, Zhu et al developed a superlens with holes in its structure to amplify evanescent waves in the near-field through resonant tunneling[157]. Such a lens could be used to improve ultrasonic imaging or detect cracks in building components. Sound absorption AMs are another popular use. Acoustic metamaterial absorbers have overlapping application with attenuation AMs; however, they are fundamentally different. Absorbers use resonators that couple with incoming waves, thus effectively trapping them with little to no reflection or transmission. They can absorb both low- and high-frequency waves depending on the design. Many of these materials are capable of >90% absorption making them excellent candidates for sound insulation or sound proofing[158-162] (**Figure 2**). Acoustic metamaterial sensors are used for pressure and sound sensing[163-165] and non-destructive evaluation[166].

## IV. THERMAL METAMATERIALS

Thermal metamaterials[167, 168] (TMs) manipulate the flow of heat. Heat transfer (i.e., conduction, convection, and radiation) is less ordered than wave propagation so achieving control is more difficult, which in turn makes the fabrication of TMs difficult. Heat naturally travels from higher temperatures to lower temperatures unless work is done on the system to reverse this process. This is a diffusive process that is different from wave propagation so coordinate invariant heat diffusion equations are required to create TMs with the transformation method. The effective parameters for TMs are thermal conductivity ($\kappa$), mass density ($\rho$), and specific heat capacity ($c_p$). In 2008, Fan et al.[34] and Chen et al.[33] discovered form-invariant thermal conduction equations (**Table 1**) and achieved control of heat flow by using extreme thermal conductivity values not seen in natural materials. All natural materials have fixed, positive thermal conductivities so achieving extreme or negative thermal conductivity gives rise to interesting heat flow phenomena. Fan et al. used transformation thermodynamics to develop a material capable of transferring heat from a lower temperature to a higher temperature in a process known as "apparent negative thermal conductivity." Soon after, Chen et al. proved that transformation thermodynamics still works in curvilinear anisotropic backgrounds and developed a cloaking device. Fan and Chen only worked with steady-state systems, but most heat phenomena involve transient systems that switch between unsteady and steady states. In 2012, Guenneau et al.[169] developed a transformation method, known as "transformation thermodynamics," for steady-state and transient systems that was later experimentally verified by Narayana and Sato[170], Narayana et al.[171], and Schittny et al.[169, 172]. In addition to TMs, PNCs also can be used to manipulate heat flow[22, 23]. Remember that phonons are units of vibrational energy, and vibrations produce sound as well as heat. This relationship between vibrations, sound, and heat allows PNCs to be used in both acoustics and thermodynamics. Similar to AMs, TMs can be constructed on the macro- or microscales depending on the method and target application. At the microscale, heat is carried by phonons that oscillate at THz frequencies, so they behave more like particles than waves. This makes it difficult to control thermal phonons with PNCs.

Despite fabrication challenges, many groups have created transformation thermodynamics-based devices such as thermal cloaks, concentrators[173], rotators[174], and camouflage[175, 176]. A thermal cloak keeps its cloaked region at a constant temperature while under a thermal gradient, thereby appearing to be thermally invisible. A thermal concentrator focuses thermal flux from a finite volume to a smaller one (**Figure 3B**). A thermal rotator can reverse the direction of heat flux within a region. Thermal camouflage has a similar result to the cloak but, instead, mimics the thermal properties of another object, effectively making the camouflaged object look like another object. Practical applications for devices like these are needed, but some groups have proposed good potential applications. Dede et al.[177] proposed using thermal cloaks, concentrators, and rotators in printed circuit boards for heat control in electronics. Also, active versions of these devices that are sensitive to voltage[178] and light[179] stimuli are being investigated. Most devices, such as those mentioned above, are meant to control thermal conduction due to the use of form-invariant conduction equations for transformation thermodynamics. Metamaterials for thermal convection[180-182] and radiation[183, 184] control currently are being studied. TMs also are being studied and fabricated on the microscopic level through phonon or photon manipulation. Thermal diodes[128, 129, 185, 186] (**Figure 2**), thermal transistors[187], thermal logic gates[188-190], and thermal memories[191] are all possible through phonon and photon manipulation.

## V. SEISMIC METAMATERIALS

Seismic metamaterials[192, 193] (SMs) have gained much attention in the past several years as an emerging metamaterial field. Elastic metamaterials always have been of interest in the acoustic and seismic fields, but SMs are tailored to low-frequency elastic waves. Seismic metamaterials manipulate the propagation of seismic waves through deflection, attenuation, and absorption. The main goals of SMs are wave attenuation and wave redirection to protect buildings and other structures from seismic waves produced by earthquakes. The effective parameters giving SMs their unique properties are mass density ($\rho$) and the Lamé parameters ($\lambda$ and $\mu$) that when negative, give rise to negative refraction. Local resonance and Bragg scattering are the two main mechanisms for inducing negative or extreme properties in SMs. Seismic wave propagation is governed by elastodynamic equations called the Navier equations. In 2006, Milton et al.[35] investigated form-invariant elastodynamic equations and discovered that the standard Navier equations are not coordinate invariant in their general form. Instead, they must be altered to Willis equations to be form invariant (**Table 1**). However, at times, the Navier equations do hold under coordinate transformation[194], and other sets of elastodynamic equations have been shown to be coordinate invariant[195]. Because of the difficulty of using coordinate-invariant equations, many groups implement local resonance and Bragg scattering through periodic arrangement of effective parameter materials. In 2012, Farhat et al.[36] proposed a cylindrical elastic cloak for elastic waves propagating through thin plates that through numerical simulation, successfully controlled propagation of the waves. This led to experiments on seismic waves carried out by Brule et al.[196, 197] in 2014 and 2017 in which they successfully demonstrated that periodically structured soils are able to attenuate seismic waves through inclusions, or holes, in the soil. In addition to SMs, PLTCs[24] and certain AMs are used for seismic wave manipulation. PLTCs are analogous to PTCs and PNCs but implement a periodic structure of thin elastic plates that manipulate vibrations. However, PLTCs are in the early stages of research so they are not as widely used as PTCs or PNCs.

Seismic metamaterials have been classified in different ways, usually based on characteristics such as wave cloaking, attenuation, and deflection. Because of the size of seismic waves (tens to hundreds of meters), many SMs are physically and spatially large enough to be able to manipulate seismic waves. Other metamaterials have small lattice constants while SMs tend to have lattice constants on the meter scale. In general, we will be classifying SMs into two main groups[192]: 1) outer shield metamaterials and 2) foundation metamaterials. Outer shields refer to the metamaterials that interact with seismic waves outside of the structure they are meant to protect, ultimately preventing the seismic waves from reaching the building. Foundation metamaterials implement metamaterial designs or devices into the foundations of buildings to combat seismic waves (**Figure 2**). Seismic soil metamaterials are an example of outer shields that implement periodic design based on transformation methods. Brule et al.[196-198] developed seismic soils consisting of cylindrical voids and rigid inclusions that interfere with seismic wave propagation. Voids and inclusions (i.e., holes) dug into the ground in a periodic, in some cases non-periodic, array cause seismic bandgaps, complete reflection, wave-path control, or attenuation by energy dissipation depending on their arrangement in the soil. Resonators buried in the soil are used to reduce the response of the soil to seismic waves. Such resonators consist of a mass, a spring, and a damper that are tuned to the structural frequency of the building. When excited, the damper resonates out of phase with the structural motion of the building, thus dissipating the energy caused by seismic waves[198]. The shape, size, and position of the resonators are directly linked to bandwidth and attenuation efficiency of the bandgaps so these qualities can vary. Some have suggested burying cross-shaped, locally resonant cylinders[199], some have tried resonant cylinders with differing eigenfrequencies[200], and others have tested cubic arrays of resonant spheres[201]. Above-surface resonators are resonators placed on the ground around a designated structure to interfere with incoming seismic waves. These resonators can be artificial (i.e., like the seismic metawedge[202]) or natural (i.e. such as trees in a forest[203, 204]). The metawedge attenuates the wave as it passes through or converts it to a much less harmful bulk shear wave. Trees can attenuate seismic waves by creating bandgaps through local resonance. Metamaterial foundations are the other classification of SMs in which the foundation of the building, or parts of it, is a metamaterial or has a metamaterial design. Many designs consist of periodic, layered sections of material like the one developed by Xiang et al.[205], which implement layers of concrete and rubber and Cheng and Shi[206] which uses steel, concrete, and rubber. Other groups [138, 207] have tested periodic plate foundations to successfully produce seismic bandgaps. Auxetic materials, discussed in further detail in the next section, can be buried in the soil underneath a building to act as a part of the foundation and create bandgaps[208]. Chiral materials (i.e., materials that cannot be superimposed upon themselves) also can be used to mitigate seismic wave activity. Carta et al.[209] proposed using gyro-elastic beams arranged in a chiral design to reduce the vibrations on a building and the concept successfully created bandgaps. Implementing metamaterials into the foundation is a promising form of SM that protects a structure from incoming waves, thus reducing the effects of displacement and stress.

In addition to the civil engineering applications of SMs, there are also long-range subsurface sensing applications. (**Figure 2**, **Figure 3C**). Porous and/or flexible metal-organic framework (MOF) composites and nanoparticles are acoustic metamaterials with tunable sound adsorption and resonances.[210] Injectable MOF suspensions have been shown to alter the distinct elastic and anelastic properties of rocks, resulting in decreased seismic wave velocities and amplitudes.[136, 211] The ability to disperse the contrast agent nanoparticles[212] throughout a fluid-rock system allows for the possibility of imposing a 2D or 3D km-scale subsurface meta-structure (**Figure 3C**). Overall, these attributes make injectable seismic contrast agents a potentially transformational technology for enabling geophysical sensing and lends new perspective to the burgeoning field of seismic metamaterials.[193, 213]

## VI. MECHANICAL METAMATERIALS

Mechanical metamaterials[2] (MMs) are another emerging field that in recent years, has received much attention because of their potential uses and ability to be implemented in other fields like AMs and SMs. Mechanical metamaterials are periodic, sometimes aperiodic, structures that use mechanical motion, energy, deformations, and stresses in unconventional ways to give these metamaterials extraordinary moveability. Initially MMs exploited geometry to achieve negative or extreme effective parameters, but recently, designs have extended beyond this. The effective parameters of MMs are mass density ($\rho$) and the elastic moduli (i.e., Young's modulus [$E$], bulk modulus [$K$], shear modulus [$G$], and Poisson's ratio [v]) (**Table 1**). When these parameters are made negative or extreme. the metamaterial possessing these qualities can exhibit extraordinary mechanical movement. Individual unit cells in MMs move together and/or in response to one another to produce behavior not seen in the constituent materials. A prime example is an auxetic MM. Auxetic materials possess a negative Poisson's ratio, giving them the ability to expand or contract in all directions when force is applied[39, 214]. Normally when a material is pulled or stretched in one direction it will become thinner in the perpendicular directions exhibiting a positive Poisson's ratio; however, this is not the case for auxetic materials that expand radially when pulled transversely. Chiral, extremal, and pentamode metamaterials are more examples of MMs. Chirality is useful for not only elastic wave manipulation but also for mechanical deformations. MMs capable of converting uniaxial compression into torsion[215] or inducing their own global rotational motion[216] are possible through chirality. Extremal metamaterials can resist one specific deformation. For example, an extremal metamaterial can greatly resist isotropic compression but will buckle under shear compression. Pentamode metamaterials[217], or metafluids, possess fluid-like properties. They are known as pentamode metamaterials because they possess five very small eigenvalues making them extremely compliant to deformation in five of six primary directions. This results in a small shear modulus and large bulk modulus when compared to one another. A large bulk modulus allows the volume of the metamaterial to remain the same under deformation that can be interpreted as a Poisson's ratio of 0.5. A small shear modulus gives the metamaterial the ability to flow like water. These elastic moduli give pentamode materials their fluid-like properties. Most pentamode metamaterials use the atomic lattice developed by Milton and Cherkaev[40] in which two conical beams connected at their base are arranged in a diamond lattice structure. This design[218] and others[219-223] have been fabricated and tested yet all still are only approximations of an ideal pentamode metamaterial. Pentamode metamaterials can be used to manipulate elastic waves in acoustics and seismology. They also are good candidates for additive manufacturing[223, 224] of metamaterials that require a complex distribution of mechanical properties.

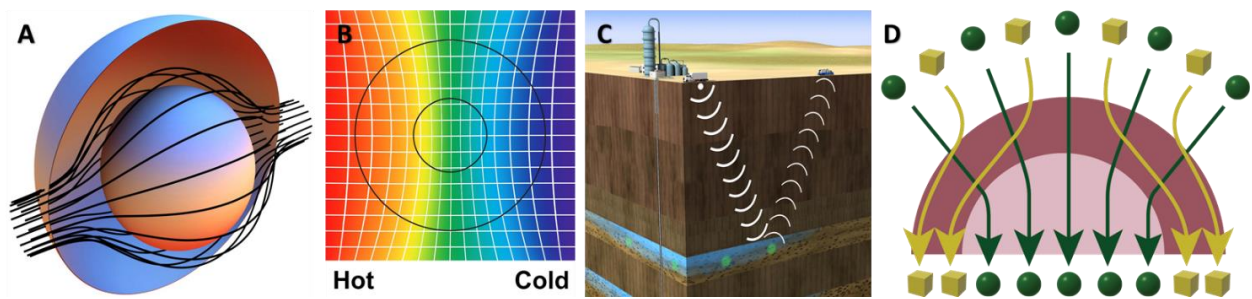

**Figure 3. A)** Optical cloak: A three-dimensional view of the optical cloak. The light rays are guided within the annulus of the cloaking material contained within R1 < r < R2 and emerge from the opposite side uninterrupted from their original course. **B)** Thermal concentrator: A constant heat source is applied to the left of the concentrator. The mesh formed by streamlines of thermal flux (vertical) and isothermal values (horizontal) illustrates the deformation of the transformed thermal space which is squeezed into the central disc. **C)** Seismic metamaterial sensing: Enhanced monitoring of subsurface fluids and structures (e.g., fractures) after injection of contrast agents that form periodic meta-structures. **D)** Mass separation: Anisotropic membrane consisting of an isotropic cylindrical core of radius R1 covered by an anisotropic cylindrical shell with internal radius R1 and external radius R2. The two molecules permeate in the membrane where one compound is directed around the core and the other compound is directed towards the core. Graphic inspired by Pendry et al.[71] (A), Guenneau et al.[169] (B), Miller et al.[136] (C), and Restrepo-Florez and Maldovan[130] (D).

Bertoldi et al.[2] breaks MMs down into three categories: 1) mechanism-based, 2) instability-based, and 3) topological metamaterials. Mechanism-based MMs are the most straightforward. A mechanism is a group of moving parts connected by hinges or some other type of linkage that perform one motion or action together. Origami and kirigami principles are commonly used in mechanism-based MMs. Origami involves precise folding that allows materials to contract, expand, and bend without breaking. Kuribayashi et al.[131] developed flexible medical stents (**Figure 2**) using this technique that can expand and contract when needed. Two-dimensional metasheets also have been fabricated[225]. Kirigami involves making precise cuts that allow the shape and elastic properties of the material to be controlled[226]. Isobe et al.[227] fabricated stretchable sheets from stiff materials. Instability-based MMs use elastic instabilities and large deformations to create strong nonlinear relationships between the stresses and strains of the MM. There are many buckling-based MMs that rely on elastic instability. A simple example is an elastomeric sheet of material possessing a square array of holes that allows the sheet to compress more than traditional material. The thin pieces of material formed from the addition of holes behave like beams buckling under compression[228]. Elastic instability also is used in snapping-based MMs that have units able to snap into certain positions based on the compression or extension of the metamaterial. Topological MMs possess topologically protected properties meaning their geometric properties remain the same though deformations such as stretching, bending, and twisting. Mechanical metamaterials with topologically protected properties have a topological index that is unchanged, which allows the materials to manipulate mechanical wave propagation and absorb energy.

## VII. MASS TRANSPORT METAMATERIALS

Mass transport metamaterials (MTMs) are an emerging field of metamaterials that can manipulate mass flow in diffusion processes. Like TMs, MTMs are designed based on diffusion laws rather than wave propagation laws. In fact, a group of researchers has theorized metamaterials capable of controlling mass or heat flow due to the mathematical similarities of their governing equations[229]. Fick's laws dictate mass diffusion and Fick's second law is invariant under coordinate transformation, which is required for using the transformation method (**Table 1**). For MTMs, the effective parameter is the chemical potential gradient of the system. Classically, the driving force for diffusion is the concentration gradient; however, for MTMs, Restrepo-Flórez and Maldovan[37, 38, 130, 230] realized it is chemical potential gradient due to the non-homogeneous nature of metamaterials. Also, MTMs usually are non-homogenous and highly anisotropic due to the spatial distribution of diffusivity required for guided mass flow. However, this area of metamaterials still is mostly unexplored because the fundamentals of mass diffusion are not well understood.

Current MTMs aim to control mass flow to assist in separation (**Figure 2**) and catalysis processes. Restrepo-Flórez and Maldovan have developed several MTMs capable of concentrating and/or cloaking mass flux. There is no standard coordinate transformation method for mass flow manipulation so multiple metamaterial configurations have been tested. For example, one device is a multilayered structure consisting of concentric rings that separate a binary mixture of oxygen and nitrogen diffusing through a polymeric matrix by cloaking nitrogen and concentrating oxygen[37]. Another separation device performs the same task but uses an anisotropic cylindrical membrane consisting of an isotropic core and an anisotropic shell[130] (**Figure 3D**). Traditionally isotropic membranes are used to separate compounds by permeating one species and rejecting the other, but anisotropic membranes allow both species to permeate while guiding them to different spatial locations.[130] However non-homogenous, anisotropic materials are extremely difficult to fabricate and manufacture so layers of homogeneous, isotropic material are used to achieve the desired anisotropic properties. Very few groups are working on MTMs[37, 38, 130, 229-236] and even fewer can successfully fabricate them because of the difficulties encountered. Improved coordinate transformation techniques for mass flow control are needed for this field to progress.

## VIII. OUTLOOK

Over the past 20 years, significant progress has been made in the design and fabrication of metamaterials. The field has expanded beyond electrodynamics to acoustics, thermodynamics, seismology, mechanics, and mass transport. Phenomena such as negative refraction, bandgaps, near perfect wave absorption, wave focusing, negative Poisson's ratio, negative thermal conductivity, etc., all are possible with these metamaterials. Improvements in metamaterial design, fabrication, and manufacturing are crucial to finding more practical applications. Electromagnetic and acoustic metamaterials are perhaps the most researched and understood materials but require better fabrication and manufacturing methods. Thermal metamaterials require further research on heat transfer and phonons. Real-world scenarios involving heat are more complicated than current TMs are capable of, and phonon transfer still is not quite fully understood. Seismic metamaterials are relatively new metamaterials with the first field experiments occurring in the 2010s. The applications are obvious and abundant for SMs; however, more field testing needs to be done to better understand their effectiveness on seismic waves. Current MMs perform novel functions, but real applications are scarce. MTMs require a better transformation technique to standardize the fabrication and realize the many possible applications for MTMs. This Research Update has thoroughly summarized the history, current state of progress, and emerging directions of metamaterials by field. Practical metamaterial devices and structures will soon be available with continued research in metamaterials.

## ACKNOWLEDGEMENTS

This material is based on work supported by the U.S. Department of Energy Office of Fossil Energy at PNNL through the National Energy Technology Laboratory, Morgantown, West Virginia. JEH was partially supported by the U.S. Department of Energy National Nuclear Security Administration MSIIP program. We thank the Editor and two anonymous reviewers for their thorough and helpful reviews.

## AUTHOR INFORMATION

JEH: 0000-0002-9094-8553
HTS: 0000-0002-4546-3979
BPM: 0000-0002-1486-5836
QRSM: 0000-0003-3009-9702

## SELECTED ABBREVIATIONS

AM; acoustic metamaterial

EM; electromagnetic metamaterial

MM; mechanical metamaterials

MTM; mass transport metamaterials

NIM; negative index materials

PTC; photonic crystal

PNC; phononic crystal

PLTC; platonic crystal

SM; seismic metamaterial

TM; thermal metamaterial

**Table 1.** The coordinate invariant governing equations for each category that are manipulated to design metamaterials.

| Field | Equation Name | Equations |
|---|---|---|
| Electromagnetic | Maxwell's Equations | $\nabla \cdot \mathbf{D} = \rho$ <br> $\nabla \cdot \mathbf{B} = 0$ <br> $\nabla \times \mathbf{E} = -\frac{\partial \mathbf{B}}{\partial t}$ <br> $\nabla \times \mathbf{H} = \frac{\partial \mathbf{D}}{\partial t} + \mathbf{J}$ |
| | | D = electric displacement, $\rho$ = charge density, B = magnetic field, E = electric field, H = magnetic field strength, J = current density |
| Acoustic | Acoustic Wave Equation | $\frac{\partial^2 p}{\partial t^2} = c^2 \nabla^2 p$ |
| | | p = pressure, c = speed of sound |
| Thermal | Heat Conduction Equation | $\rho C \frac{\partial T}{\partial t} + \nabla(-\kappa \nabla T) = Q$ |
| | | $\rho$ = density, C = heat capacity, T = temperature, $\kappa$ = thermal conductivity, Q = heat source |
| Seismic | Elastodynamic Equations (Willis Equations) | $\text{div } \sigma = \dot{p}$ <br> $\sigma = C^{eff} * e + S^{eff} * \dot{u}$ <br> $p = S^{eff} * e + \rho^{eff} * \dot{u}$ <br> $e = \frac{1}{2}(\nabla u + (\nabla u)^t)$ |
| | | $\sigma$ = stress, e = strain, p = momentum density <br> $C^{eff}$, $S^{eff}$, & $\rho^{eff}$ = non-local operators |
| Mechanical | Elastic Moduli (Young's Modulus, Bulk Modulus, Shear Modulus) Poisson's Ratio | $E = \frac{\sigma}{\varepsilon} = \frac{\text{stress}}{\text{strain}}$ <br> $K = \frac{p_1 - p_0}{V_1 - V_0/V_0} = \frac{\text{volumetric stress}}{\text{volumetric strain}}$ <br> $G = \frac{\tau_{xy}}{\gamma_{xy}} = \frac{\text{shear stress}}{\text{shear strain}}$ <br> $\nu = -\frac{\varepsilon_t}{\varepsilon_l} = \frac{\text{transverse strain}}{\text{longitudial strain}}$ |
| Mass Transport | Diffusion Equation (Fick's 2nd Law) | $\frac{\partial c_i}{\partial t} = \nabla(D_i \nabla c_i)$ |
| | | $c_i$ = concentration, $D_i$ = diffusion coefficient |

90. Alibakhshikenari, M., et al., *Metamaterial-Inspired Antenna Array for Application in Microwave Breast Imaging Systems for Tumor Detection.* IEEE Access, 2020. **8**: p. 174667-174678.
91. AlSabbagh, H.M., et al., *A compact triple-band metamaterial-inspired antenna for wearable applications.* Microwave and Optical Technology Letters, 2020. **62**(2): p. 763-777.
92. Li, L.W., et al., *A broadband and high-gain metamaterial microstrip antenna.* Applied Physics Letters, 2010. **96**(16).
93. Watts, C.M., X.L. Liu, and W.J. Padilla, *Metamaterial Electromagnetic Wave Absorbers.* Advanced Materials, 2012. **24**(23): p. Op98-Op120.
94. Landy, N.I., et al., *Perfect metamaterial absorber.* Physical Review Letters, 2008. **100**(20): p. 207402.
95. Tao, H., et al., *A metamaterial absorber for the terahertz regime: design, fabrication and characterization.* Optics express, 2008. **16**(10): p. 7181-7188.
96. Tao, H., et al., *A dual band terahertz metamaterial absorber.* Journal of physics D: Applied physics, 2010. **43**(22): p. 225102.
97. Zhu, J., et al., *Ultra-broadband terahertz metamaterial absorber.* Applied Physics Letters, 2014. **105**(2): p. 021102.
98. Avitzour, Y., Y.A. Urzhumov, and G. Shvets, *Wide-angle infrared absorber based on a negative-index plasmonic metamaterial.* Physical Review B, 2009. **79**(4): p. 045131.
99. Liu, X., et al., *Infrared spatial and frequency selective metamaterial with near-unity absorbance.* Physical Review Letters, 2010. **104**(20): p. 207403.
100. Aydin, K., et al., *Broadband polarization-independent resonant light absorption using ultrathin plasmonic super absorbers.* Nature communications, 2011. **2**(1): p. 1-7.
101. Jiang, W., et al., *Electromagnetic wave absorption and compressive behavior of a three-dimensional metamaterial absorber based on 3D printed honeycomb.* Scientific Reports, 2018. **8**.
102. Yang, J.J., et al., *Metamaterial Sensors.* International Journal of Antennas and Propagation, 2013. **2013**.
103. Withayachumnankul, W., et al., *Metamaterial-based microfluidic sensor for dielectric characterization.* Sensors and Actuators a-Physical, 2013. **189**: p. 233-237.
104. Ebrahimi, A., et al., *High-Sensitivity Metamaterial-Inspired Sensor for Microfluidic Dielectric Characterization.* Ieee Sensors Journal, 2014. **14**(5): p. 1345-1351.
105. Zhang, Y., et al., *Microwave metamaterial absorber for non-destructive sensing applications of grain.* Sensors, 2018. **18**(6): p. 1912.
106. Cong, L., et al., *Experimental demonstration of ultrasensitive sensing with terahertz metamaterial absorbers: A comparison with the metasurfaces.* Applied Physics Letters, 2015. **106**(3): p. 031107.
107. Saadeldin, A.S., et al., *Highly sensitive terahertz metamaterial sensor.* IEEE Sensors Journal, 2019. **19**(18): p. 7993-7999.
108. Chen, T., S. Li, and H. Sun, *Metamaterials application in sensing.* Sensors, 2012. **12**(3): p. 2742-2765.
109. Xiao, S.Y., et al., *Active metamaterials and metadevices: a review.* Journal of Physics D-Applied Physics, 2020. **53**(50).
110. Pitchappa, P., et al., *Bidirectional reconfiguration and thermal tuning of microcantilever metamaterial device operating from 77 K to 400 K.* Applied Physics Letters, 2017. **111**(26): p. 261101.
111. Tao, H., et al., *MEMS based structurally tunable metamaterials at terahertz frequencies.* Journal of Infrared, Millimeter, and Terahertz Waves, 2011. **32**(5): p. 580-595.
112. Zhao, X., et al., *Voltage-tunable dual-layer terahertz metamaterials.* Microsystems & Nanoengineering, 2016. **2**(1): p. 1-8.
113. Karvounis, A., et al., *Nano-optomechanical nonlinear dielectric metamaterials.* Applied Physics Letters, 2015. **107**(19): p. 191110.
114. Zhang, J., K.F. MacDonald, and N.I. Zheludev, *Nonlinear dielectric optomechanical metamaterials.* Light: Science & Applications, 2013. **2**(8): p. e96-e96.
115. Shadrivov, I.V., et al., *Metamaterials Controlled with Light.* Physical Review Letters, 2012. **109**(8): p. 083902.
116. Zhang, X.G., et al., *Light-Controllable Digital Coding Metasurfaces.* Advanced Science, 2018. **5**(11): p. 1801028.
117. Zhang, X.G., W.X. Jiang, and T.J. Cui, *Frequency-dependent transmission-type digital coding metasurface controlled by light intensity.* Applied Physics Letters, 2018. **113**(9): p. 091601.
Page 17 of 22